\documentclass[10pt, conference, a4]{IEEEtran}
\IEEEoverridecommandlockouts
\usepackage{cite}
\usepackage{amsmath,amssymb,amsfonts}
\usepackage{algorithmic}
\usepackage{graphicx}
\usepackage{textcomp}
\usepackage{xcolor}
\def\BibTeX{{\rm B\kern-.05em{\sc i\kern-.025em b}\kern-.08em
    T\kern-.1667em\lower.7ex\hbox{E}\kern-.125emX}}

\usepackage{cite}
\usepackage[ansinew]{inputenc}
\usepackage{psfrag,color}
\usepackage[labelformat=simple]{subcaption}
\usepackage{amsfonts,amssymb,amsthm,amscd,amsmath}
\usepackage{mathtools}
\usepackage{commath,mathabx}
\usepackage{enumerate}
\usepackage{algorithmic}
\usepackage[export]{adjustbox}
\usepackage{textcomp}
\usepackage{xcolor}
\usepackage[overload]{empheq}

\newcommand{\T}{{\scalebox{.65}{$\rm T$}}}

\newcommand{\E}{{\rm E}}
\newcommand{\uM}{{\mathbf u}}

\newcommand{\w}{{\mathbf w}}

\definecolor{laranja}{rgb}{0.8,0.5,0}
\definecolor{capri}{rgb}{0.0, 0.75, 1.0}
\definecolor{carmine}{rgb}{0.59, 0.0, 0.09}
\definecolor{dimgray}{rgb}{0.41, 0.41, 0.41}

\newcommand{\boxedeqn}[1]{
	\begin{equation}
	\fbox{$\displaystyle #1 $}
	\end{equation}}

\usepackage{fancyhdr}

\fancypagestyle{plain}{
	\fancyhf{}
	\fancyhead[C]{\textcolor{blue}{Paper accepted for publication in the Proc. of the 28th European Signal Processing Conference (EUSIPCO), 18-22 January, 2021, Amsterdam, NL.}}     
	\fancyfoot[L]{This is a notice}

}
\usepackage{eso-pic}

	\AddToShipoutPictureBG*{%
		\AtPageUpperLeft{%
			\setlength\unitlength{1in}%
			\hspace*{\dimexpr0.5\paperwidth\relax}
			\makebox(0,-0.5)[c]{\Large \textcolor{blue}{Paper accepted for publication in the Proc. of the 28th European Signal}}%
			
			\makebox(0,-1)[c]{ \Large \textcolor{blue}{Processing Conference (EUSIPCO), 18-22 January, 2021, Amsterdam, NL.}}%
	}}
	\AddToShipoutPictureBG*{%
		\AtPageLowerLeft{%
			\setlength\unitlength{1in}%
			\hspace*{\dimexpr0.5\paperwidth\relax}
	}}

\title{A Sampling Algorithm for Diffusion Networks}

\author{Daniel G.~Tiglea, Renato Candido, and Magno T.~M.~Silva\\
	Escola Politécnica, University of São Paulo, Brazil\\
	\{dtiglea, renatocan, magno\}@lps.usp.br
\vspace*{-0.2cm}	\thanks{This work was supported by FAPESP under Grant 2017/20378-9, by CNPq under Grants 132586/2018-5 and 304715/2017-4, and by CAPES under Finance Code 001.}}

\begin{document}

\abovedisplayskip=2pt
\belowdisplayskip=2pt
\abovedisplayshortskip=0pt
\belowdisplayshortskip=2pt
\thispagestyle{fancy}

	%
	\maketitle
	\begin{abstract}
		In this paper, we propose a sampling mechanism for adaptive diffusion networks that adaptively changes  the amount of sampled nodes based on
		mean-squared error in the neighborhood of each node. It presents fast convergence during transient and a significant reduction in the number of sampled nodes in steady state.
		Besides reducing the computational cost, the proposed mechanism can also be used as a censoring technique, thus saving energy by reducing the amount of communication between nodes.
		We also present a theoretical analysis to obtain lower and upper bounds for the number of network nodes sampled in steady state.
	\end{abstract}
	
\vspace*{-0.1cm}
	\section{Introduction}
\vspace*{-0.1cm}
	Over the last decade, adaptive diffusion networks
	have attracted widespread attention  since they can be used
	to efficiently estimate certain parameters of interest using information  collected at spatially distributed nodes  connected through a particular
	topology \cite{Sayed_Networks2014,lopes2008diffusion,cattivelli2009diffusion,takahashi2010diffusion,yu2013strategy,Fernandez-Bes2017,lopes2008topologies,zhao2012single,xu2015adaptive,takahashi2010link,arroyo2013censoring,fernandez2015censoring}.
	Many efforts have been devoted to obtain diffusion strategies
	that are able to learn and adapt from continuous streaming data
	and exhibit fast convergence, good tracking capability, and low computational cost.
	When these strategies are implemented on wireless sensor networks,
	energy consumption is the most critical constraint~\cite{takahashi2010link, arroyo2013censoring,fernandez2015censoring}.


As a result, several selective transmission mechanisms have been proposed to reduce the energy consumption associated with the communication processes. Some of these approaches aim to reduce the amount of information sent in each transmission~\cite{arablouei2013distributed,chouvardas2013trading}, while others turn links off according to selective communication policies~\cite{lopes2008topologies,takahashi2010link,zhao2012single,xu2015adaptive}. Finally, a certain set of solutions seeks to censor the nodes by avoiding the transmission of information to any of their neighbors~\cite{arroyo2013censoring,fernandez2015censoring,berberidis2015adaptive,yang2018distributed}.  This allows the censored nodes  to turn their transmitters off, thus saving more energy, and reduces the amount of information used in the processing~\cite{fernandez2015censoring,berberidis2015adaptive}.

	
	Recently, we proposed in \cite{Tiglea_Eusipco2019} a sampling mechanism for the graph diffusion algorithm of
	\cite{NassifICASSP2018}. This mechanism changes adaptively the amount of sampled nodes in the graph based on
	mean-squared error (MSE) in the neighborhood of each node. Thus, the number of sampled nodes decreases when the MSE is low, allowing for fast convergence in the transient and a significant reduction in the computational cost in steady state.
	
	This paper extends our previous work \cite{Tiglea_Eusipco2019} in different
	ways:
	(i) the algorithm of \cite{Tiglea_Eusipco2019} is generalized to adaptive diffusion networks in order to
	reduce their computational cost,
	(ii) we obtain theoretical lower and upper bounds for the number of network sampled nodes in steady state,
	and (iii) we show that, with slight modifications, the proposed scheme can also be used as a censoring technique.
	
	The paper is organized as follows. In Sec.~\ref{problem_formulation}, we revisit the Adapt-Then-Combine diffusion Normalized Least-Mean-Squares (ATC dNLMS) algorithm~\cite{Sayed_Networks2014}. In Sec.~\ref{sampling}, the adaptive sampling algorithm is derived. In Sec.~\ref{sec:analysis}, we present a theoretical analysis to predict
	bounds for the number of sampled nodes in steady state.
	Simulation results are shown in Sec.~\ref{experiments} and Sec.~\ref{conclusions} closes the paper with the conclusions.

	\noindent\textbf{Notation}. We use normal fonts for scalars and boldface letters for vectors. Moreover, $(\cdot)^{\T}$ denotes transposition, $|\cdot|$ cardinality, $\E\{\cdot \}$ the mathematical expectation and $\lVert\cdot\rVert$ the Euclidean norm.
	
	\vspace*{-0.1cm}
	\section{Distributed Adaptive Filtering} \label{problem_formulation}
	\vspace*{-0.1cm}
	Let us consider a network of $V$ nodes with a predefined topology. Two nodes are considered neighbors if they can exchange information, and we denote by $\mathcal{N}_k$ the neighborhood of node $k$ including $k$ itself.
	Each node $k$ has access to an input signal $u_k(n)$ and to a reference signal $d_k(n) = \uM_k^{\T}(n)\mathbf{w}^{\mathrm{o}} + v_k(n)$, where $\mathbf{u}_k(n) = [u_k(n)\,  \ u_k(n\!-\!1)\, \cdots\, u_k(n\!-\!M\!+\!1)]^{\T}$
	is an $M$-length regressor vector, $\mathbf{w}^{\mathrm{o}}$ is an optimal system, and $v_k(n)$ is the measurement noise at node $k$, which is assumed to be independent of the other variables and zero-mean
	with variance $\sigma^2_{v_k}$. The objective of the network is to obtain an estimate of $\mathbf{w}^\mathrm{o}$ in a distributed manner by solving
	\mbox{$\min_{\mathbf{w}} \sum_{k=1}^V \E \{ \lvert d_k(n) - \uM_k^{\T}(n) \mathrm{w} \rvert^2 \}$} \cite{Sayed_Networks2014,lopes2008diffusion, cattivelli2009diffusion,takahashi2010link, arroyo2013censoring}.
	
	Several adaptive solutions have been proposed in the literature for this task, one of them being the ATC dNLMS algorithm~\cite{Sayed_Networks2014,lopes2008diffusion, cattivelli2009diffusion}. It consists in two steps, and its equations are given by
	\begin{subequations} \label{eq:lms_sayed}
		\begin{empheq}[left={\empheqlbrace\,}]{align}
		&\boldsymbol{\psi}_k(n+1)\!=\!\w_k(n)\!+\!\mu_k(n) \uM_k(n) e_k(n) \label{eq:lms_sayed1}\\
		&\w_k(n+1)\!=\!\textstyle\sum_{j \in \mathcal{N}_k} c_{jk} \boldsymbol{\psi}_j(n+1), \label{eq:lms_sayed2}
		\end{empheq}
	\end{subequations}
	where \vspace*{-0.1cm}
	\begin{equation} \label{eq:ek}
	e_k(n) = d_k(n) - \uM_k^{\T}(n) \w_k(n),
	\end{equation}
	$\boldsymbol{\psi}_k$ and  $\w_k$ represent respectively the estimation error and the local and combined estimates of $\w^{\rm o}$ at node $k$,
	and $\mu_k(n) = {\widetilde{\mu}_k}/{[\delta + \|\uM_k(n)\|^2]}$ is a normalized step size with $0<\!\!\widetilde{\mu}_k\!<\!2$ and $\delta\!\!>\!\!0$ a
	small constant~\cite{Sayed_Networks2014}. Furthermore, $\{c_{jk}\}$ are combination weights satisfying
	$c_{jk}\!\geq\!0$, $\sum_{j\in \mathcal{N}_k}\!c_{jk}\!=\!1$, and $c_{jk}\!=\!0$ for $j\notin \mathcal{N}_k$~\cite{lopes2008diffusion, cattivelli2009diffusion}.
	Possible choices for $\{c_{jk}\}$ include the Uniform, Laplacian,
	Metropolis, and Relative Degree rules~\cite{Sayed_Networks2014}, as well as adaptive
	schemes~\cite{Fernandez-Bes2017,takahashi2010diffusion,yu2013strategy}, such as the Adaptive Combination Weights (ACW) algorithm~\cite{tu2011optimal}. It incorporates information from the noise profile across the network, and is obtained by solving an optimization problem in regards to $\{c_{jk}\}$. It can be summarized as~\cite{tu2011optimal}
	\begin{equation} \label{eq:acw1}
	c_{jk}(n)=\frac{\sigma^{-2}_{jk}(n)}{\sum_{\ell \in \mathcal{N}_k}\sigma^{-2}_{\ell k}(n)}\ \text{if}\ j \in \mathcal{N}_k\;\;\text{or}\;\; 	0, \text{otherwise},
	\end{equation}
	where $\sigma^{2}_{jk}$ is updated as
	\begin{equation} \label{eq:acw2}
	\sigma^2_{jk}(n)\! =\! (1\!-\!\nu_k)\sigma^2_{jk}(n\!-\!1)\!+\!\nu_k \lVert\boldsymbol{\psi}_j(n\!+\!1)\!-\!\w_k(n)\rVert^2,
	\end{equation}
	with $\nu_k\!>\!0$ for $k\!=\!1,\cdots,V$. Hence, greater weights are assigned to the nodes with smaller noise variances~\cite{tu2011optimal}.
	
It is worth noting that one could also employ a Combine-Then-Adapt (CTA) strategy, in which the order of~\eqref{eq:lms_sayed1} and~\eqref{eq:lms_sayed2} is reversed~\cite{Sayed_Networks2014}. For simplicity, in this paper we only consider the ATC strategy in our analysis, but the results can be straightforwardly extended to CTA versions as well.
	
	\vspace*{-0.3cm}
	\section{The sampling algorithm}\label{sampling}
	\vspace*{-0.2cm}
	We propose an algorithm to decide if each node of the network should be sampled or not at each iteration. For this purpose, we introduce the variable $\widebar{s}_k(n) \! \in \! \{0,1\}$ and \mbox{recast~\eqref{eq:lms_sayed1} as}
	\vspace*{-0.1cm}
	\boxedeqn{ \label{eq:lms_sayed_mod}
		\boldsymbol{\psi}_k(n+1) = \w_k(n) + \widebar{s}_k(n) \mu_k(n)  \uM_k(n) e_k(n).
	}
	If $\widebar{s}_k(n)\!=\!1$, $d_k(n)$ is sampled, $e_k(n)$ is computed as in \eqref{eq:ek}, the combination weights are updated according to~\eqref{eq:acw1} and~\eqref{eq:acw2}, and \eqref{eq:lms_sayed_mod} coincides with \eqref{eq:lms_sayed1}. In contrast, if $\widebar{s}_k(n)\!=\!0$, $d_k(n)$ is not sampled, $\uM_k^{\T}(n) \w_k(n)$, $e_k(n)$ and $\mu_k(n)$ are not computed, the $\{c_{jk}\}$ are not updated and \mbox{$\boldsymbol{\psi}_k(n\!+\!1)\!=\!\w_k(n)$}. 
	
	To determine $\widebar{s}_k(n)$, we define $s_k(n) \! \in \! [0,1]$ such that $\widebar{s}_k(n)\!=\!0$ for $s_k(n)\!<\! 0.5$ and $\widebar{s}_k(n)\!=\!1$ otherwise.
	We then minimize the following cost function with respect to $s_k(n)$:
	\begin{equation} \label{eq:cost}
	\!\!J_{s,k}(n)\! =\! [s_k(n)]\beta \bar{s}_k(n)\!+\!\left[1\!-\!s_k(n)\right] \textstyle\sum_{j \in \mathcal{N}_k}\!\!\! c_{ik}(n)e_i^2(n),
	\end{equation}
	where  $\beta \!\! > \!\! 0$ is a parameter introduced to control how much the sampling of the nodes is penalized. Thus, when the error is high in magnitude or when node $k$ is not being sampled ($\bar{s}_k\!=\!0$), $J_{s,k}(n)$ is minimized by making $s_k(n)$ closer to one, leading to the sampling of node $k$. This ensures that the algorithm keeps sampling the nodes while the error is high and resumes the sampling of
	idle nodes at some point, enabling it to detect changes in the environment. In contrast, when node $k$ is being sampled ($\bar{s}_k\!=\!1$) and the error is small in magnitude in comparison to $\beta$, $J_{s,k}(n)$ is
	minimized by making $s_k(n)$ closer to zero, which leads the algorithm to stop sampling node $k$. This desirable behavior depends on a proper choice for $\beta$, which is addressed in Sec.~\ref{sec:analysis}. 
	
	Inspired by convex combination of adaptive filters (see~\cite{SPM2016,Garcia_Biased2010} and their references),
	rather than directly adjusting $s_k(n)$, we update an auxiliary
	variable $\alpha_k(n)$ related to it via \cite{Garcia_Biased2010}
	\begin{equation} \label{eq:tb}
	s_k(n) = \phi[\alpha_k(n)] \triangleq \frac{\mathrm{sgm}[\alpha_k(n)]-\mathrm{sgm}[-\alpha^+]}{\mathrm{sgm}[\alpha^+]-\mathrm{sgm}[-\alpha^+]},
	\end{equation}
	where $\mathrm{sgm}[x]\!=\!(1\!+\!e^{-x})^{-1}$
	is a sigmoidal function and $\alpha^{\! +}$ is the maximum value $\alpha_k$ can assume.
	We should notice that $\phi[\alpha^{\! +}\!] \! = \! 1$ and $\phi[-\alpha^{\! +}\!]\!=\!0$. In the literature, $\alpha^{\! +}\!=\!4$ is usually adopted~\cite{Garcia_Biased2010}.
	
	By taking the derivative of \eqref{eq:cost} with respect to $\alpha_k(n)$, we obtain
	the following stochastic gradient descendent rule:
	\begin{equation} \label{eq:b_original}
	\!\!\alpha_k(\!n\!+\!1\!)\!=\!\alpha_k(\!n\!)\!+\!\mu_s\phi^{\prime}[\alpha_k(\!n\!)]\!\! \left[\!\textstyle\sum_{i \in \mathcal{N}_k}\!\!\!c_{ik}\!(\!n\!) e_i^2(\!n\!)\!-\!\beta \bar{s}_k(\!n\!) \! \right]\!\!,\!
	\end{equation}
	where $\mu_s>0$ is a step size and
	\begin{equation} \label{eq:phi_linha}
	\phi^{\prime}[\alpha_k(n)]\!\triangleq\!\frac{d s_k(n) }{d\alpha_k(n)}\!=\! \frac{\mathrm{sgm}[\alpha_k(n)]\{1\!-\!\mathrm{sgm}[\alpha_k(n)]\}}{\mathrm{sgm}[\alpha^+]\!-\!\mathrm{sgm}[-\alpha^+]}.
	\end{equation}

	Equation~\eqref{eq:b_original} cannot be used for sampling since it requires the errors to be computed  to decide if the nodes should be sampled or not,
	which is contradictory. To address this issue, we replace $e_i(n)$ in~\eqref{eq:b_original} by its latest measurement we have access to,
	which is denoted by $\varepsilon_i(n)$. When the node is sampled, $\varepsilon_i(n)\!=\!e_i(n)$. We thus obtain 
	\boxedeqn{ \label{eq:b_modificada}
		\!\alpha_k(\!n\!+\!1\!)\!=\!\alpha_k(\!n\!)\!+\!\mu_s\phi^{\prime}[\!\alpha_k(\!n\!)\!]\! \Big[\!\!\!\!\sum_{\;\; i \in \mathcal{N}_k}\!\!\!\! c_{ik}(\!n\!) \varepsilon_i^2(\!n\!)\!-\!\beta  \widebar{s}_k(\!n\!) \! \Big]\!.
	}
	
	This algorithm is named as adaptive sampling diffusion NLMS (AS-dNLMS).
	It reduces the number of sampled nodes in steady state,  decreasing the computational cost
	at the expense of a slight increase  during the transient. Table~\ref{tab:cost} shows
	the number of sums and multiplications executed per iteration in a single node of the network for both the
	dNLMS and AS-dNLMS algorithms with ACW weights. When the node is sampled, AS-dNLMS requires $\sum_{i \in \mathcal{N}_k}\!\bar{s}_i(n)\!+\!2$
	more multiplications and $|\mathcal{N}_k|\!+\!1$ more additions than the original dNLMS algorithm.
	On the other hand, when the node is not sampled, AS-dNLMS requires $3M\!+\!2\!-\!\sum_{i \in \mathcal{N}_k}\!\bar{s}_i(n)$
	less multiplications and $4M\!-\!|\mathcal{N}_k|+1$ less sums. Thus, the higher the order of the filter $M$, the higher
	the computational cost reduction of AS-dNLMS in comparison with the original dNLMS algorithm. Considering the network as a whole,
	the computational cost of AS-dNLMS depends on the number of sampled nodes, which is addressed in Sec.~\ref{sec:analysis}.

	Finally, we remark that an alternate version of AS-dNLMS can be obtained if, instead of using~\eqref{eq:lms_sayed_mod}, $\boldsymbol{\psi}_k$
	is not updated at all when node $k$ is not sampled.
	Assuming that the nodes can store past information from their neighbors,
	this allows us to cut the number of communications between nodes, since in this case $\boldsymbol{\psi}_k$ and $\varepsilon^2_k$
	remain static when $\bar{s}_k\!=\!0$ and there is no need for node $k$ to retransmit them. In other words, when node $k$ is not
	sampled in this version of the algorithm, it only receives data and carries out~\eqref{eq:lms_sayed2}, and can thus turn its transmitter off.
	This results in a reduction in energy consumption as well as the  computational~cost.
	Lastly, when the node is sampled, $\varepsilon_i^2(n)=e_i^2(n)$ can be
	sent bundled with the local estimates
	$\boldsymbol{\psi}_i$ in both versions of AS-dNLMS so as to not increase the number of transmissions.

	\begin{table*}[t]
		\centering
		\caption{\label{tab:cost} Comparison between dNLMS and AS-dNLMS: number of operations per iteration for each node $k$.}
		\vspace*{-0.1cm}
		\begin{tabular}{|c|c|c|}
			\hline
			\hline
			Algorithm & Multiplications ($\bigotimes$) & Sums ($\bigoplus$) \\
			\hline
			dNLMS & $M(3+|\mathcal{N}_k|)+4$ & $M(3+|\mathcal{N}_k|)+3$ \\
			\hline
			AS-dNLMS & $\bar{s}_k(n) \cdot(3M+4)+M|\mathcal{N}_k|+\sum_{i \in \mathcal{N}_k}\!\bar{s}_i(n)+2$ & $\bar{s}_k(n)\cdot(4M+2)+M|\mathcal{N}_k|-M+|\mathcal{N}_k|+2$\\
			\hline
			\hline
		\end{tabular}
		\vspace*{-0.4cm}
	\end{table*}

	%
	
	\vspace*{-0.3cm}
	\section{Theoretical analysis} \label{sec:analysis}
	\vspace*{-0.2cm}
	The good behavior of AS-dNLMS depends on a proper choice for $\beta$.
	Thus, we study how to choose this parameter such that we can ensure that every node 
will cease to be sampled at some point during steady state. To do so, we examine~\eqref{eq:b_modificada} while node $k$ is being sampled. 
In this case, $\varepsilon^2_i(n)$ and $\beta \bar{s}_k(n)$ can be replaced by $e^2_i(n)$ and $\beta$, respectively. 
Then, subtracting $\alpha_k(n)$ from both sides of \eqref{eq:b_modificada} and taking expectations, we get
	\begin{equation} \label{eq:expected_b}
	\E\{\Delta \alpha_k(\!n\!) \}\!=\!\mu_s  \E \Big\{\! \phi^{\prime}[\alpha_k(\!n\!)]\! \left[  \textstyle\sum_{i \in \mathcal{N}_k}\!\!c_{ik}(\!n\!)e_i^2(\!n\!)\!-\!\beta \right]\!\Big\}.
	\end{equation}
	where  $\Delta \alpha_k(n) \triangleq \alpha_k(n\!+\!1)\!-\!\alpha_k(n)$.
	To make the analysis more tractable,
	$\phi^{\prime}[\alpha_k(n)]$ and the term between brackets in~\eqref{eq:expected_b} are  assumed statistically. Although this may seem unrealistic, simulation results suggest it is a reasonable approximation.
	Thus, in order to stop sampling node $k$,
	$\alpha_k(n)$ should decrease along the iterations until it becomes negative.
	Since $\phi^{\prime}[\alpha_k(n)]$ is always positive, to
	enforce $\E\{\Delta \alpha_k(n)\}$ to be negative while node $k$ is sampled, $\beta$ must
	satisfy \vspace*{-0.1cm}
	\begin{equation} \label{eq:beta1}
	\beta > \textstyle\sum_{i \in \mathcal{N}_k}\!\!c_{ik}(n) \E\{e_i^2(n)\}.
	\end{equation}
	Assuming that the order of the adaptive filter is sufficient and that $\tilde{\mu}_k$, $k\!=\!1,2,\cdots\!,V,$ are chosen properly so that the gradient noise can be disregarded, it is reasonable to assume that in steady state $\E\{e_i^2(n)\} \approx \sigma_{v_{i}}^2$, which leads to
	\begin{equation}
	\label{eq:menorque}
	\textstyle\sum_{i \in \mathcal{N}_k}\!\!c_{ik}(n) \E \{e_i^2(n) \} \leq \sigma^2_{\max} \triangleq \max_{i} \sigma_{v_{i}}^2,
	\end{equation}
	where $i=1,2,\cdots,V$.
	Thus, the  condition 
	\begin{equation} \label{eq:beta2}
	\beta > \sigma^2_{\max}
	\end{equation}
	is  sufficient to ensure
	that, in the mean, the nodes will cease to be sampled during steady state.
	
	Assuming that \eqref{eq:beta2} is satisfied, we can estimate upper and lower bounds for the expected number of sampled nodes  $V_s$ in steady state. For this purpose, we consider each $\widebar{s}_k(n)$ as a Bernoulli random variable that is equal to one with probability $p_{s_k}$ or to zero with probability $1-p_{s_k}$ in steady state for $k=1, \ \cdots, \ V$. Thus, \vspace*{-0.1cm}
	\begin{equation} \label{eq:bounds}
	V p_{s_{\min}} \leq  \E\{V_s\} \leq V p_{s_{\max}},
	\end{equation}
	where $p_{s_{\min}}$ and $p_{s_{\max}}$ are upper and lower bounds for $p_{s_k}$.

	It is useful to note that the sampling mechanism exhibits a cyclic behavior in steady state. Hence, we could approximate $p_{s_k}$ by the expected ``duty cycle'' of the mechanism, i.e.,
	\begin{equation} \label{eq:psk}
	\displaystyle \widehat{p}_{s_k} = {\theta_k}/{(\theta_k+\overline{\theta}_k)},
	\end{equation}
	where $\theta_k$ denotes the expected number of iterations per cycle in which node $k$ is sampled and $\overline{\theta}_k$ is the expected number of iterations in which it is not.
	Since we are only interested in estimating $p_{s_{\min}}$ and $p_{s_{\max}}$, we do not have to evaluate~\eqref{eq:psk} for every $k$.
	Instead, we only need to estimate upper and lower bounds for $\theta_k$~and~$\overline{\theta}_k$. To do so, we must understand under which circumstances node $k$ remains sampled for the greatest (or lowest) number of iterations in the mean. One way to do this is to estimate the maximum and minimum values $\E\{\alpha_k(n)\}$ and $\E\{\Delta \alpha_k(n)\}$ can assume during steady state.

	Firstly, let us assume that at a certain iteration $n$, $\alpha_k(n)$ is negative but close to zero.
	Thus, setting $\alpha_k(n)$ to zero in~\eqref{eq:b_modificada} and taking expectations, we obtain
	\begin{equation} \label{eq:limite1}
	\E\{\alpha_k(n+1)\} = \mu_s \phi_0^\prime \textstyle\sum_{i \in \mathcal{N}_k}\!\!c_{ik}(n) \E \{\varepsilon_i^2(n) \},
	\end{equation}
	where $\phi_0^\prime\!\!=\!\!\phi^\prime[0]$. Thus, at $n\!+\!1$ the sampling of node $k$ resumes and,
	recalling~\eqref{eq:beta2}, $\E\{\Delta \alpha_k(n\!+\!1)\}\!<\!0$. Therefore, from iteration $n\!+\!1$ onwards,
	$\alpha_k$ decreases until it becomes negative again, meaning that~\eqref{eq:limite1} yields the maximum value $\alpha_k$ can assume in the mean during steady state. Moreover, assuming $\sigma^2_{\min}\! \leq\! \E \{\varepsilon_i^2(n) \} \! \leq\! \sigma^2_{\max}$ for all $i$,~\eqref{eq:limite1} yields a different value for each node $k$ that lies in
	\begin{equation}
	\mu_s \phi_0^\prime \sigma^2_{\min} \leq \E\{\alpha^{\text{s.s.}}_{k_{\max}}\} \leq \mu_s \phi_0^\prime \sigma^2_{\max},
	\end{equation}
	where $\E\{\alpha^{\text{s.s.}}_{k_{\max}}\}$ denotes the maximum value $\alpha_k(n)$ can assume in the mean in steady state and $\sigma^2_{\min}\!\triangleq\! \min_{i} \sigma_{v_{i}}^2$, $i\!=\!1,\cdots,V$. Analogously, we now assume that at a certain iteration $n$, $\alpha_k(n)$ is positive but approximately zero. Making this replacement in~\eqref{eq:b_modificada} and taking expectations, we obtain
	\begin{equation} \label{eq:limite2}
	\E\{\alpha_k(n+1)\} = \mu_s \phi_0^\prime \E \left\{\textstyle\sum_{i \in \mathcal{N}_k}\!\!c_{ik}(n) \varepsilon_i^2(n) -\beta \right\}.
	\end{equation}
	Since $\alpha_k(n)<0$ and $\E\{\Delta \alpha_k(n)\}>0$ while node $k$ is not being sampled,~\eqref{eq:limite2} provides the minimum value $\alpha_k$ can assume in the mean during steady state. For each node $k$,~\eqref{eq:limite2} yields a different value that lies in the interval
	\begin{equation}
	\mu_s \phi_0^\prime (\sigma^2_{\min}-\beta) \leq \E\{\alpha^{\text{s.s.}}_{k_{\min}}\} \leq \mu_s \phi_0^\prime (\sigma^2_{\max}-\beta),
	\end{equation}
	where $\E\{\alpha^{\text{s.s.}}_{k_{\min}}\}$ denotes the minimum value $\alpha_k(n)$ can assume in the mean in steady state. 

	Next, we replace $\phi^\prime[\alpha_k(n)]$ in~\eqref{eq:b_modificada} by its first-order Taylor expansion around $\alpha_k(n)\!=\!0$, which is simply equal to the constant $\phi^\prime_0$. 
	When node $k$ is being sampled ($\widebar{s}_k(n) \!=\!1$),
	subtracting $\alpha_k(n)$ from both sides of \eqref{eq:b_modificada} and taking expectations yields
	\begin{equation} \label{eq:delta_alpha1}
	\!\!-\mu_s \phi_0^\prime   (\beta \!-\! \sigma^2_{\min} ) \! \leq \!  \E\{\Delta \alpha_k(n)\} \! \leq \!  -\mu_s \phi_0^\prime   (\beta \!-\! \sigma^2_{\max} )\!<\!\!0.
	\end{equation}
	Analogously, when the node is not sampled ($\widebar{s}_k(n) =0$),
	\begin{equation} \label{eq:delta_alpha2}
	\mu_s \phi_0^\prime  \sigma^2_{\min} \! \leq \!  \E\{\Delta \alpha_k(n)\} \! \leq \!  \mu_s \phi_0^\prime \sigma^2_{\max}.
	\end{equation}
	Thus, in both cases there are upper and lower bounds for $\E\{\Delta \alpha_k(n)\}$ during steady state.
	
	From a certain iteration $n_0$ onward, we consider the model
	\begin{equation}
	\E\{\alpha_k(n_0+\theta_{k})\}\!=\!\E\{\alpha_k(n_0)\}+\theta_{k}\E\{\Delta\alpha_k(n)\}.\label{eq:geral}
	\end{equation}
	In order to estimate an upper bound $\theta_{\max}$ for $\theta_k$, we assume that
	$\E\{\alpha_k(n_0)\}=\E\{\alpha^{\text{s.s.}}_{k_{\max}}\}$ and calculate the expected number of iterations required for $\E\{\alpha_k(n)\}$ to fall
	below zero in the
	scenario where the node is sampled for the maximum number of iterations.
	This occurs if  $\E\{\alpha_k(n_0)\}\! =\! \mu_s \phi_0^\prime \sigma^2_{\max}$, which is the upper bound for
	$\E\{\alpha^{\text{s.s.}}_{k_{\max}}\}$, and $\E\{\Delta \alpha_k(n)\}\! =\! -\mu_s \phi_0^\prime (\beta - \sigma^2_{\max})$,
	which is the least negative variation for $\E\{\Delta\alpha_k(n)\}$ according to~\eqref{eq:delta_alpha1}.
	Making $\theta_k\!=\!\theta_{\max}$ and setting $\E\{\alpha_k(n_0+\theta_{\max})\}\!=\!0$ in \eqref{eq:geral}, after some algebraic manipulations we obtain
	\begin{equation} \label{eq:tmax}
	\theta_{\max} = \max\{{\sigma^2_{\max}}/{(\beta - \sigma^2_{\max})},\;\; 1 \},
	\end{equation}
	where we are taking into account the fact that the node must be sampled at least once during each cycle. Analogously,
	using \eqref{eq:geral} for the lower bound $\theta_k=\theta_{\min}$, we obtain
	\begin{equation} \label{eq:tmin}
	\theta_{\min} = \max\{{\sigma^2_{\min}}/{(\beta - \sigma^2_{\min})},\;\; 1 \}.
	\end{equation}
	For $\overline{\theta}_k$, we replace $\theta_k$ in \eqref{eq:geral} by $\overline{\theta}_k$ and consider that at the iteration $n_0$,
	$\E\{\alpha_k(n_0)\}\!=\!\E\{\alpha^{\text{s.s.}}_{k_{\min}}\}$.
	Thus, the upper bound $\overline{\theta}_{\max}$ for $\overline{\theta}_k$
	can be obtained by setting
	$\E\{\alpha_k(n_0)\}\! =\! \mu_s \phi_0^\prime \sigma^2_{\min}$,
	which is the lower bound for $\E\{\alpha^{\text{s.s.}}_{k_{\min}}\}$, and $\E\{\Delta \alpha_k(n)\}\! =\! \mu_s \phi_0^\prime \sigma^2_{\min}$,
	which is the minimum value for $\E\{\Delta\alpha_k(n)\}$ according to~\eqref{eq:delta_alpha2}.
	We then get
	\begin{equation} \label{eq:t_barramax}
	\overline{\theta}_{\max} = \max\{{(\beta - \sigma^2_{\min})}/{\sigma^2_{\min}},\;\; 1 \}.
	\end{equation}
	Analogously, for the lower bound $\overline{\theta}_{\min}$ of $\overline{\theta}_k$, we get	
	\begin{equation} \label{eq:t_barramin}
	\overline{\theta}_{\min} = \max\{{(\beta - \sigma^2_{\max})}/{\sigma^2_{\max}},\;\; 1 \}.
	\end{equation}
	Replacing~\eqref{eq:tmax} to~\eqref{eq:t_barramin} in~\eqref{eq:psk} and~\eqref{eq:bounds}, after some algebraic manipulations we finally obtain
	\boxedeqn{
		\label{eq:final}
	V\dfrac{\sigma_{\min}^2}{\beta} \leq \E\{V_s\} \leq V\dfrac{\sigma_{\max}^2}{\beta}.}
	
	This indicates that the higher the parameter $\beta$, the smaller the amount of sampled nodes
	in the mean during steady state, which is in accordance with our expectations.
	Since there is a trade-off between the tracking capability and the gains in terms of computational cost provided by the sampling mechanism,
	we should care not to  choose excessively high values for $\beta$. Simulation results suggest that
	$\beta\!>\!10\sigma_{\max}^2$ can deteriorate the performance in non-stationary environments.
	It is also worth noting that the upper and lower bounds for $\E\{V_s\}$ coincide when $\sigma^2_{\min}\! =\! \sigma^2_{\max}$. This makes sense, since in this case there is no reason for some nodes to be sampled more often than the others in steady state. Furthermore, although we initially assumed $\beta\!>\!\sigma_{\max}^2$, it is interesting to note that~\eqref{eq:final} also holds for $\beta\!=\!\sigma_{\max}^2$, since in this case the theoretical upper bound for $\E\{V_s\}$ is equal to the total number of nodes in the network. Finally, we should notice that the step size $\mu_s$ does not affect the amount of sampled nodes.

\vspace*{-0.1cm}
	\section{Simulation Results} \label{experiments}
\begin{figure}[htb]
		\centering
		\begin{subfigure}[c]{0.2\textwidth}
			\centering
			\vskip 10pt
			\includegraphics*[trim= 7cm 10.5cm 7cm 10cm,clip=true,width=0.9\columnwidth]{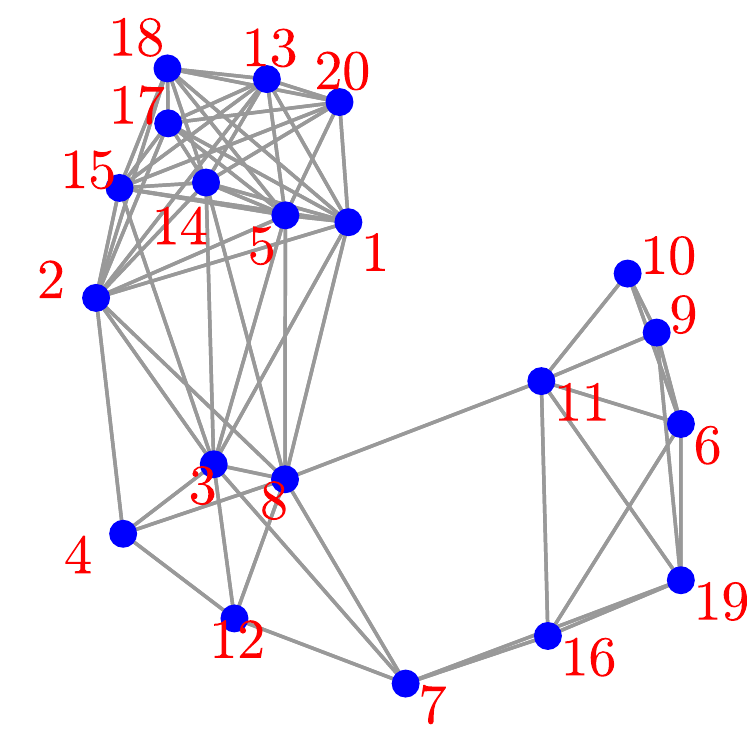}
			\caption{\label{graph_ex}}
		\end{subfigure}
		\begin{subfigure}[c]{0.24\textwidth}
			\centering
			\includegraphics*[width=1\columnwidth]{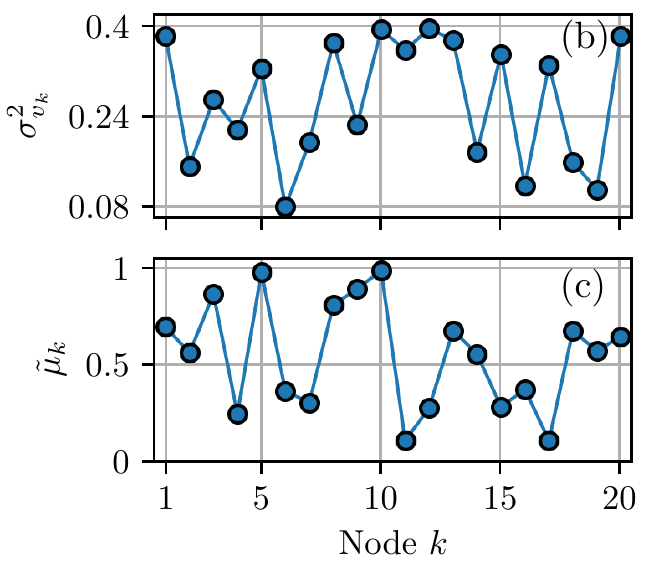}
		\end{subfigure}
		\vspace*{-0.2cm}
		\caption{\label{gsnr} \footnotesize \subref{graph_ex} Network topology, (b)~$\sigma^2_{v_k}$, and (c)~$\tilde{\mu}_k$  used in the experiments..}
		\vspace*{-0.6cm}
	\end{figure}

\vspace*{-0.1cm}
	In this section, we test the proposed algorithm and the analysis of Sec.~\ref{sec:analysis}. The results presented were obtained over an average of 100 realizations. For the sake of better visualization, we filtered the curves by a moving-average filter with $64$ coefficients. 
We consider the network shown in Fig.~\ref{gsnr}(a). The signals $u_k(n)$ and $v_k(n)$ are generated from i.i.d. zero-mean Gaussian random processes with variances $\sigma_{u{_k}}^2\!=\!1$ and  $\sigma_{v_{k}}^2$ as shown in Fig.~\ref{gsnr}(b) for $k=1,\cdots,V$. For the optimal system $\w^{\rm o}$, we consider a random vector with $M\!=\!50$ coefficients uniformly distributed in $[-1,1]$. 

\begin{figure}[htb]
		\vspace*{-0.3cm}
		\centering
		\includegraphics*[width=0.49\textwidth]{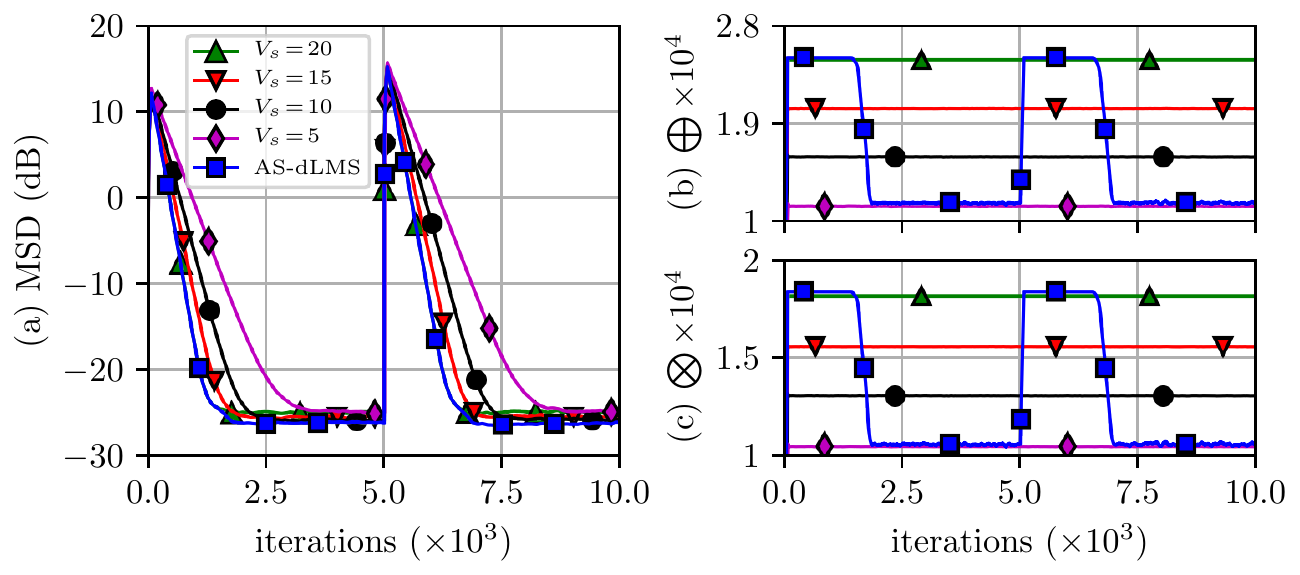}
		\vspace*{-0.7cm}
		\caption{\label{fig:M50} \footnotesize Comparison between dNLMS with  a random sampling technique with different amounts of sampled nodes and AS-dNLMS (\mbox{$\beta \!=\! 0.68$}, \mbox{$\mu_s \! = \! 0.1571$}). (a) MSD curves, (b) Sums, and (c) Multiplications per iteration.}
		\vspace*{-0.2cm}
	\end{figure}

\begin{figure}[htb]
		\vspace*{-0.3cm}
		\centering
		\includegraphics*[width=1.0\columnwidth]{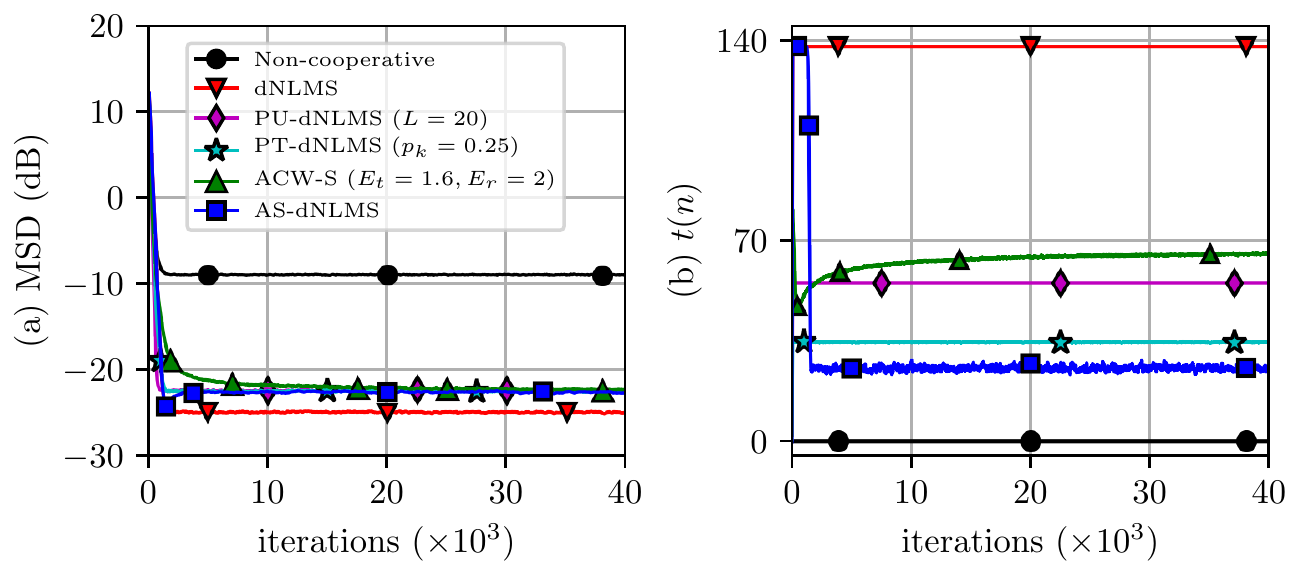}
		\vspace*{-0.5cm}
		\caption{\label{fig:transmi} \footnotesize Comparison between the energy-saving version of AS-dNLMS (\mbox{$\beta \!=\! 0.68$}, \mbox{$\mu_s \! = \! 0.1571$}) and other techniques found in the literature~\cite{arablouei2013distributed,lopes2008topologies,arroyo2013censoring}. (a) MSD, and (b) Communications per iteration.}
		\vspace*{-0.4cm}
	\end{figure}

	To set the combination weights, we use the ACW algorithm with $\nu_k\!=\!0.2$ for $k\!=\!1,\cdots,V$~\cite{tu2011optimal}. We use $\delta\!=\!10^{-5}$ and different values of $\tilde{\mu}_k$ for each node $k$, as shown in~Fig.~\ref{gsnr}(c). As a performance indicator, we adopt the network mean-square deviation (MSD), given by
	$\frac{1}{V} \sum_{k=1}^V \E \{ \norm{\w^{\rm o}(n) \!-\! \w_k(n)}^2\}$. Furthermore, in the simulations of Figs.~\ref{fig:M50} and~\ref{fig:transmi} we consider $\beta\!=\!1.7\sigma_{\max}^2\!=\!0.68$ and $\mu_s\!=\!0.1571$ for the AS-dNLMS algorithm. These values were chosen due to the good performance they provided in terms of MSD, computational cost reduction and energy saving in these simulations.
	
	Firstly, we compare the behavior of the AS-dNLMS algorithm with that of the original dNLMS with a random sampling technique
	in which $V_s$ nodes are randomly sampled at each iteration. In order to simulate a change in the environment,
	in the middle of each realization we flip $\w^{\rm o}$.
	Figs.~\ref{fig:M50}(a),~\ref{fig:M50}(b) and~\ref{fig:M50}(c)
	present respectively the MSD performance and the average number of sums and multiplications per iteration.
	We can observe from Fig.~\ref{fig:M50}(a) that the more nodes are sampled, the faster the convergence rate.
	AS-dNLMS is able to detect the change in the optimal system and, since all nodes are sampled during the transients, it converges
	as fast as the dNLMS algorithm with all nodes sampled.
	From Figs.~\ref{fig:M50}(b) and~\ref{fig:M50}(c) we also observe that during the transients the computational cost of AS-dNLMS is
	slightly higher than that of the dNLMS algorithm with all nodes sampled, but decreases significantly after AS-dNLMS
	converges and ceases to sample every node at every iteration.

\begin{figure}[htb]
		\vspace*{-0.3cm}
		\centering
		\includegraphics*[width=0.7\columnwidth]{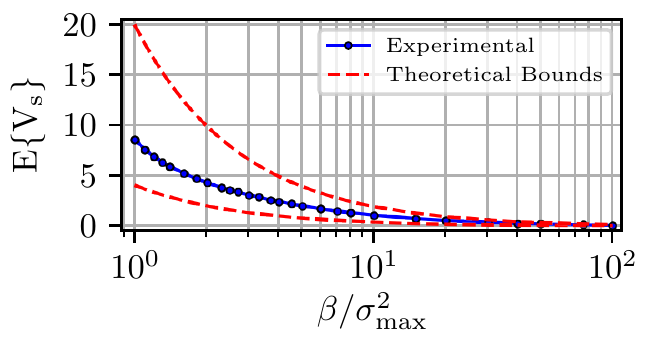}
		\vspace*{-0.3cm}
		\caption{\label{fig:beta} \footnotesize Average number of sampled nodes during steady state and comparison with theoretical bounds of~\eqref{eq:final}.}
		\vspace*{-0.3cm}
	\end{figure}

	In Fig.~\ref{fig:beta} we present simulation results showing the average number of sampled nodes
	during steady state in a stationary environment, as well as the theoretical bounds given by~\eqref{eq:final} for different
	values of $\beta/\sigma^2_{\max}\! \geq \! 1$.
	We can see that the higher $\beta$ is, the less nodes are sampled, as expected.
	Furthermore, the experimental results
	lie between the theoretical curves for all values of $\beta/\sigma^2_{\max}$, validating the results of
	Sec.~\ref{sec:analysis}. 

Finally, we consider the energy-saving version of AS-dNLMS in which node $k$ does not communicate with its neighbors when it is not sampled.
To assess its performance, we compare it with the ACW-Selective algorithm of~\cite{arroyo2013censoring} (ACW-S), the partial-update algorithm of~\cite{arablouei2013distributed} (PU-dNLMS), and the dNLMS algorithm with a probabilistic transmission strategy in which each link of the network is active at a certain iteration $n$ with probability $p_k$ (PT-dNLMS). In our simulations, we adjusted the parameters of all the algorithms to obtain roughly the same level of MSD during steady state. For comparison, we also present the results obtained with the original dNLMS algorithm and with the non-cooperative case. In Fig.~\ref{fig:transmi}(a) we present the MSD performance, and
in Fig.~\ref{fig:transmi}(b) the number of communication processes per iteration $t(n)$. To enable the comparison with the PU-dNLMS algorithm, we scaled the number of communication processes by the ratio of data sent in each transmission in this plot. We observe that AS-dNLMS initially requires just as many transmissions as the original dNLMS algorithm, but this number drastically decreases after it converges. During steady state, it led to the lowest number of communication processes among all the solutions tested. It is interesting to note that, in a scenario where the nodes are able to broadcast their estimates to all
their neighbors at once, the comparison shown in Fig.~\ref{fig:transmi}(b) is unfair with the AS-dNLMS and ACW-S algorithms, since in this case they are the only solutions that would lead to an actual reduction in the number of communication processes.

	
	\section{Conclusions} \label{conclusions}
	In this paper, we generalize the sampling mechanism of~\cite{Tiglea_Eusipco2019} for adaptive diffusion networks. The proposed mechanism uses the information from more nodes when the error in the network is high and less nodes otherwise.
	Besides reducing the computational cost, it can be used to save energy by avoiding transmissions of nodes that are not sampled.
	We observed from simulations that AS-dNLMS maintains the convergence rate of dNLMS during transient while displaying a lower computational cost in steady state.
	The energy-saving version of AS-dNLMS presents a slight increase in steady-state MSD, but still exhibits a good tradeoff between performance and energy consumption.
	The theoretical bounds for the number of sampled nodes obtained in Sec.~\ref{sec:analysis} present a good agreement with simulations, and are useful for the proper choice of algorithm parameters. It should be mentioned that, although we compared the proposed AS-dNLMS algorithm with other techniques in Sec.~\ref{experiments}, it may be used in conjunction with these methods, as well as many others~\cite{arablouei2013distributed,chouvardas2013trading,xu2015adaptive,arroyo2013censoring,fernandez2015censoring}, to further reduce the computational cost and the energy consumption associated with the communication processes.
	For future work, we intend to compare AS-dNLMS with other state-of-the-art censoring mechanisms \cite{fernandez2015censoring}
	and test it with other diffusion schemes, such as decoupled algorithms~\cite{Fernandez-Bes2017}.
	
	

\end{document}